\documentclass[english,aps,reprint,amsmath,amssymb,groupedaddress]{revtex4-2}
\usepackage[T1]{fontenc}
\usepackage[utf8]{inputenc}
\usepackage{babel}
\usepackage{amsmath}
\usepackage{graphicx}
\usepackage{txfonts}

\DeclareMathOperator{\sign}{sgn}
\DeclareMathOperator{\modulo}{mod}

\begin{document}

\title{The formation of magnetic reentrancy in the Ising model on a decorated square lattice}

\author{A. V. Zarubin}
\email{Alexander.Zarubin@imp.uran.ru}

\author{F. A. Kassan-Ogly}

\address{M. N. Mikheev Institute of Metal Physics of Ural Branch of Russian Academy of Sciences, S. Kovalevskoy Street 18, 620108 Ekaterinburg, Russia}

\begin{abstract}
In our work, based on an exact solution of the Ising model on a decorated square lattice with an arbitrary number of decorating spins, we have demonstrated the fundamental possibility of describing the phenomenon of multiple magnetic phase transitions (magnetic reentrancy) in the regime of competing exchange interactions. We analyzed the magnetic behavior of the system and established the conditions for the occurrence of magnetic reentrancy. We have determined the relationships of the model parameter under which the formation of one, three, and even five magnetic phase transitions is possible, which is confirmed by a complicated magnetic phase diagram. We have also proposed a unique method for finding critical temperatures that allows for the precise determination of their number and values, including those at extremely low temperatures.
\end{abstract}

\maketitle

\section*{Introduction}

The phenomenon of magnetic reentrancy, defined as a sequence of transitions between magnetically ordered and paramagnetic states as temperature changes, is a complicated process that has been the subject of considerable interest for many years \citep{Senoussi:1985,Mydosh:1993,Diep:2020}. This phenomenon has been observed through experimental means in a variety of magnetic systems, including spin glasses \citep{Binder:1986,Mydosh:1993}, magnetic liquids \citep{Makarov:2012}, amorphous magnets \citep{Kakehashi:1991,Kakehashi:1993}, magnetic semiconductors \citep{Calderon:2007}, perovskites \citep{Sanyal:2017,Naveen:2018}, pyrochlore \citep{Scheie:2017,Yahne:2021}, Heusler alloys \citep{Samanta:2018}, superconductors \citep{Buzdin:2005,Tran:2024}, and intermetallic compounds \citep{Chakraborty:2022}. A thorough investigation and exacting description of this phenomenon are imperative for a comprehensive understanding of the nature of multiple phase transitions in complicated magnetic systems. Such a study also has the potential to yield new insights that could lead to the development of innovative magnetic materials with unique properties.

A variety of theoretical methods have been employed to describe this phenomenon. The Ising model serves as a fundamental tool for the description and understanding of multiple magnetic phase transitions \citep{Kitatani:1985,Azaria:1987,Yokota:1989,Diep:1992,Thomas:2011,Diep:2020:1,Lajko:2020}. It is imperative to recall the fundamental principle underlying the Onsager's solution for the Ising model on an undecorated square lattice \citep{Onsager:1944,Kaufman:1949:2,Kaufman:1949:3}. This solution posits that, irrespective of the signs and values of the exchange interactions, the Ising system invariably undergoes a single magnetic phase transition. The presence of decorations in such a system has been demonstrated to either preserve or augment the number of magnetic phase transitions \citep{Kassan-Ogly:2025}. It is also noteworthy that the Ising model on a decorated square lattice in \citep{Syozi:1968,Miyazima:1968:1,Miyazima:1968:3,Nakano:1968,Nakano:1968:e,Fradkin:1976,Jascur:1998,Boughrara:2008,Strecka:2012:s,Doria:2014,Chen:2023} yielded highly intriguing results; however, no exact solution or comprehensive explanation of the observed phenomenon was attained, particularly in the context of an arbitrary decoration multiplicity of lattice.

In this paper, we will investigate the characteristics of magnetic reentrancy in the classical Ising model on a decorated square lattice with an arbitrary number of decorating spins. This investigation is founded on the exact solution \citep{Kassan-Ogly:2025}.

\section{Spontaneous magnetization of the spin system}

The spontaneous magnetization in Ising magnets is defined \citep{Montroll:1963,Tanaka:2002,Baxter:2011} as the square root of the pairwise spin-spin correlation function when the spins are spread to infinity 
\[
M=\sqrt{\lim_{\Delta\to\infty}\langle\sigma_{1,1}\sigma_{1,1+\Delta}\rangle}.
\]
The exact analytical expression for the spontaneous magnetization in the Ising model on an undecorated square lattice was first presented by Onsager in 1949 \citep{Baxter:2011} in the form of
\begin{equation}
M=\left(1-k^{2}\right)^{1/8},\quad k=\left(\sinh\frac{2J_{x}}{T}\sinh\frac{2J_{y}}{T}\right)^{-1}.\label{eq:M:O}
\end{equation}

\begin{figure}[tbh]
\centering \includegraphics{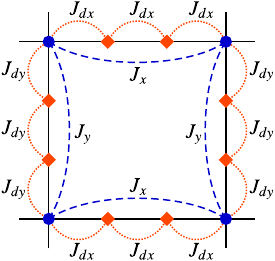}
\caption{Example of a decorated square lattice: a lattice decorated twice in two directions ($d_{x}=d_{y}=2$). Blue circles represent nodal spins; red diamonds represent decorating spins. Blue dashed lines represent direct exchange interactions with corresponding values $J_{x}$ and $J_{y}$ along the $x$ and $y$ directions of the lattice; red dotted lines represent decorating exchange interactions with corresponding values $J_{dx}$ and $J_{dy}$. The black solid lines show the spatial lattice on which the atomic spins of the lattice are located}
\label{fig:01}
\end{figure}

By generalizing this expression of Onzager (\ref{eq:M:O}) and using the concept of a decoration-iteration transformation proposed in the work of Siozi~\citep{Syozi:1951}, we obtained \citep{Kassan-Ogly:2020,Kassan-Ogly:2025} an expression for the spontaneous magnetization of the Ising model on a decorated square lattice with an arbitrary number of decorating spins in the form 
\begin{equation}
M=\left(1-k^{2}\right)^{1/8},\quad k=\left(\frac{S_{x}S_{y}}{D_{x}D_{y}}\right)^{-1},\label{eq:M:D}
\end{equation}
where
\begin{multline}
S_{i}=\frac{1}{2}e^{2\frac{J_{i}}{T}}\left(\cosh^{d_{i}+1}\frac{J_{di}}{T}+\sinh^{d_{i}+1}\frac{J_{di}}{T}\right)^{2}\\
-\frac{1}{2}e^{-2\frac{J_{i}}{T}}\left(\cosh^{d_{i}+1}\frac{J_{di}}{T}-\sinh^{d_{i}+1}\frac{J_{di}}{T}\right)^{2},\label{eq:Si}
\end{multline}
\begin{equation}
D_{i}=\left(\cosh^{d_{i}+1}\frac{J_{di}}{T}\right)^{2}-\left(\sinh^{d_{i}+1}\frac{J_{di}}{T}\right)^{2},\quad i=x,\ y.\label{eq:Di}
\end{equation}
In the functions presented, $J_{i}$ denotes the exchange interaction between atomic spins at the nearest-neighbor nodal sites of the initial lattice in the $i$ direction (direct exchange interaction); $J_{di}$ is the exchange interaction both between neighboring decorating spins and between neighboring decorating and nodal lattice spins in the $i$ direction (decorating exchange interaction); $d_{i}$ is the numbers of decorating spins (decoration multiplicities), representing the number of decorating spins located between the nodal spins in the $i$ direction; here, the index $i$ takes the values $x$ or $y$, depending on the lattice direction along which the corresponding exchange interaction acts; $T$ is absolute temperature. An example of such a decorated square lattice is shown in Fig.~\ref{fig:01}.

It is imperative to acknowledge that the Boltzmann constant has been assumed to be a unity in the presented calculations, as conventionally used in low-dimensional physics. The values of temperature and exchange interaction parameters are quantitated in absolute units of direct exchange interaction along the $x$ direction of the lattice to ensure a uniform scale.

In the following examination, the magnetic behavior of an Ising spin system on a decorated square lattice will be investigated under model parameter interrelations that yield multiple magnetic phase transitions, otherwise known as the phenomenon of magnetic reentrancy.

\section{Configuration diversity of the model}

An investigation of the magnetic properties of the model under consideration on a decorated square lattice, based on an exact solution, is a multi-parameter problem with two parameters of direct exchange interaction ($J_{x}$ and $J_{y}$), two parameters of decorating exchange interaction ($J_{dx}$ and $J_{dy}$), and two parameters of decoration multiplicity ($d_{x}$ and $d_{y}$). The exchange interaction parameters are capable of assuming any continuous value and can be negative. In this scenario, positive exchange interaction parameters ($J>0$) are indicative of a ferromagnetic type of interaction between spins, while a negative sign ($J<0$) corresponds to an antiferromagnetic type of interaction between spins. Additionally, the decoration multiplicity are regarded as discrete variables that correspond to the infinite set of natural numbers ($d>0$, $d\in\mathbb{Z}$).

It is clear that at low temperatures, along one of the directions of the square lattice---that is, along the lines of direct and decorating exchange interactions---and depending on the signs of the exchange interaction parameters, either a ferromagnetic (F) or antiferromagnetic (AF) ordering of the lattice spins may arise. Combinations of these magnetic orders along two lattice directions will yield a complicated picture of magnetism in a given decorated spin system, which manifests itself in different or identical types of spin order along different lattice directions. This is precisely why, for the accurate identification and differentiation of magnetic phases, a nomenclature is used that reflects the sequence of ordering types along the lattice directions: first along the $x$ direction, then along the $y$ direction. This approach allows for the systematization and unambiguous characterization of magnetic structures, taking into account their spatial orientation.

Consequently, the magnetic phase diagram may comprise four magnetically ordered phases: ferromagnetic-ferromagnetic (F--F) (i.e., ferromagnetic ordering along the $x$ axis of the lattice and ferromagnetic ordering along the $y$ axis of the lattice), ferromagnetic-antiferromagnetic (F--AF), antiferromagnetic-ferromagnetic (AF--F), and antiferromagnetic-antiferromagnetic (AF--AF), as well as a paramagnetic phase (P).

In the field of phase transition theory, it is widely recognized that the logarithmic divergence (a lambda-peak-type singularity) observed by Onsager \citep{Onsager:1944} in the temperature dependence of the heat capacity at critical temperatures is directly associated with the divergence of the integrable logarithm of the Helmholtz free energy (per spin)
\begin{equation}
F=-T\ln\lambda\label{eq:F}
\end{equation}
at the following four points in the integration region 
\begin{equation}
\{(0,0),\ (0,\pi),\ (\pi,0),\ (\pi,\pi)\}.\label{eq:Lf:a}
\end{equation}
Here, $\lambda$ is the principal (according to the Frobenius--Perron theorem, the only one maximal positive real) eigenvalue of the Kramers--Wannier transfer matrix \citep{Kramers:1941:1,Kramers:1941:2,Baxter:1982}.

In this case, the analytical expression for the principal eigenvalue of the Kramers--Wannier transfer matrix for a decorated square lattice with an arbitrary number of decorating spins \citep{Kassan-Ogly:2025} can be expressed as 
\begin{equation}
\ln\frac{\lambda}{2}=\frac{1}{8\pi^{2}(1+d_{x}+d_{y})}\intop_{0}^{2\pi}\intop_{0}^{2\pi}L(\phi_{x},\phi_{y})\,d\phi_{x}\,d\phi_{y},\label{eq:L1}
\end{equation}
\begin{equation}
L(\phi_{x},\phi_{y})=\ln(C_{x}C_{y}-S_{x}D_{y}\cos\phi_{x}-D_{x}S_{y}\cos\phi_{y}),\label{eq:Lf}
\end{equation}
where 
\begin{multline}
C_{i}=\frac{1}{2}e^{2\frac{J_{i}}{T}}\left(\cosh^{d_{i}+1}\frac{J_{di}}{T}+\sinh^{d_{i}+1}\frac{J_{di}}{T}\right)^{2}\\
+\frac{1}{2}e^{-2\frac{J_{i}}{T}}\left(\cosh^{d_{i}+1}\frac{J_{di}}{T}-\sinh^{d_{i}+1}\frac{J_{di}}{T}\right)^{2},\quad i=x,\ y.\label{eq:Ci}
\end{multline}
The other functions are defined above by the expressions (\ref{eq:Si}) and (\ref{eq:Di}).

This expression (\ref{eq:L1}) is predicated on Onzager's solution \citep{Onsager:1944,Kaufman:1949:2,Kaufman:1949:3,Baxter:1982,Pathria:2021} for the Ising model on a square lattice and the concept of the decoration-iterative transformation proposed in Syozi's work \citep{Syozi:1951}.

Determining the characteristics of the temperature dependence of the function (\ref{eq:Lf}) for the variables $(\phi_{x},\phi_{y})$ at the points (\ref{eq:Lf:a}) allows us to establish not only the number and values of the temperatures of magnetic phase transitions, but also to determine the type of the system primary magnetic order within the specified intervals of the model parameters. In particular, an analysis of the behavior of the function (\ref{eq:Lf}) at the point $(\phi_{x},\phi_{y})=(0,0)$ reveals the state of the system in the ferromagnetic ordering of the system in both the $x$ and $y$ directions. At the point $(\phi_{x},\phi_{y})=(\pi,\pi)$, this function demonstrates the state of the system under antiferromagnetic ordering also in both the $x$ and $y$ directions. At the point $(\phi_{x},\phi_{y})=(0,\pi)$, this function shows the state of the system with ferromagnetic ordering in the $x$ direction and antiferromagnetic ordering in the $y$ direction. At the point $(\phi_{x},\phi_{y})=(\pi,0)$, this function shows the state of the system with antiferromagnetic ordering in the $x$ direction and ferromagnetic ordering in the $y$ direction.

Therefore, by employing the expressions for the function (\ref{eq:Lf}) at the designated points (\ref{eq:Lf:a}), we can derive equations that delineate the boundaries of the corresponding magnetic phases within the magnetic phase diagram of the spin system. In particular, the equation employed to determine the phase boundaries of the region with F--F ordering is as follows 
\begin{multline}
\frac{T}{2}\ln\left(-\frac{\sinh\frac{J_{y}}{T}\sinh^{d_{y}+1}\frac{J_{dy}}{T}+\cosh\frac{J_{y}}{T}\cosh^{d_{y}+1}\frac{J_{dy}}{T}}{\sinh\frac{J_{y}}{T}\cosh^{d_{y}+1}\frac{J_{dy}}{T}+\cosh\frac{J_{y}}{T}\sinh^{d_{y}+1}\frac{J_{dy}}{T}}\right.\\
\times\left.\frac{\sinh^{d_{x}+1}\frac{J_{dx}}{T}-\cosh^{d_{x}+1}\frac{J_{dx}}{T}}{\sinh^{d_{x}+1}\frac{J_{dx}}{T}+\cosh^{d_{x}+1}\frac{J_{dx}}{T}}\right)=J_{x},\label{eq:PB:FF}
\end{multline}
with F--AF ordering --
\begin{multline}
\frac{T}{2}\ln\left(\frac{\sinh\frac{J_{y}}{T}\sinh^{d_{y}+1}\frac{J_{dy}}{T}+\cosh\frac{J_{y}}{T}\cosh^{d_{y}+1}\frac{J_{dy}}{T}}{\sinh\frac{J_{y}}{T}\cosh^{d_{y}+1}\frac{J_{dy}}{T}+\cosh\frac{J_{y}}{T}\sinh^{d_{y}+1}\frac{J_{dy}}{T}}\right.\\
\times\left.\frac{\sinh^{d_{x}+1}\frac{J_{dx}}{T}-\cosh^{d_{x}+1}\frac{J_{dx}}{T}}{\sinh^{d_{x}+1}\frac{J_{dx}}{T}+\cosh^{d_{x}+1}\frac{J_{dx}}{T}}\right)=J_{x},\label{eq:PB:FA}
\end{multline}
with AF--F ordering --
\begin{multline}
\frac{T}{2}\ln\left(-\frac{\sinh\frac{J_{y}}{T}\cosh^{d_{y}+1}\frac{J_{dy}}{T}+\cosh\frac{J_{y}}{T}\sinh^{d_{y}+1}\frac{J_{dy}}{T}}{\sinh\frac{J_{y}}{T}\sinh^{d_{y}+1}\frac{J_{dy}}{T}+\cosh\frac{J_{y}}{T}\cosh^{d_{y}+1}\frac{J_{dy}}{T}}\right.\\
\times\left.\frac{\sinh^{d_{x}+1}\frac{J_{dx}}{T}-\cosh^{d_{x}+1}\frac{J_{dx}}{T}}{\sinh^{d_{x}+1}\frac{J_{dx}}{T}+\cosh^{d_{x}+1}\frac{J_{dx}}{T}}\right)=J_{x},\label{eq:PB:AF}
\end{multline}
with AF--AF ordering --
\begin{multline}
\frac{T}{2}\ln\left(\frac{\sinh\frac{J_{y}}{T}\cosh^{d_{y}+1}\frac{J_{dy}}{T}+\cosh\frac{J_{y}}{T}\sinh^{d_{y}+1}\frac{J_{dy}}{T}}{\sinh\frac{J_{y}}{T}\sinh^{d_{y}+1}\frac{J_{dy}}{T}+\cosh\frac{J_{y}}{T}\cosh^{d_{y}+1}\frac{J_{dy}}{T}}\right.\\
\times\left.\frac{\sinh^{d_{x}+1}\frac{J_{dx}}{T}-\cosh^{d_{x}+1}\frac{J_{dx}}{T}}{\sinh^{d_{x}+1}\frac{J_{dx}}{T}+\cosh^{d_{x}+1}\frac{J_{dx}}{T}}\right)=J_{x}.\label{eq:PB:AA}
\end{multline}
It is important to acknowledge that these equations facilitate the construction of a magnetic phase diagram for the spin system under consideration, thereby illustrating the relationship between any two model parameters.

\section{Competing exchange interactions}
\label{sec:conc}

It has been established that on an undecorated square lattice at all variants of exchange interactions signs the competition is impossible. However, in a decorated square lattice where a second type of exchange interaction---the decorative interaction---is present, the competition between exchange interactions may arise. It should be noted that such competition does not occur across the entire range of model parameter interrelations; rather, it manifests itself exclusively in the cases listed below.

For odd values of the decoration multiplicity ($d_{i}$) along the $i$ axis ($i=x,y$), the competition between exchange interactions arises when the model parameters satisfy the condition
\begin{equation}
J_{i}<0,\quad|J_{di}|>0\quad d_{i}=2k_{i}+1,\quad k_{i}\in\mathbb{Z},\label{eq:CJ:o}
\end{equation}
that is, when the direct exchange interaction is negative (antiferromagnetic) and the sign of the decorating exchange interaction along the $i$ axis is arbitrary.

For even values of the decoration multiplicity ($d_{i}$), the competition between exchange interactions arises within the following range of model parameter relationships
\begin{equation}
J_{i}J_{di}<0,\quad d_{i}=2k_{i},\quad k_{i}\in\mathbb{Z},\label{eq:CJ:e}
\end{equation}
that is, when the direct and decorative exchange interactions have opposite signs along the $i$ axis. In all other cases, no competition between exchange interactions occurs in the system.

In our previous work \citep{Kassan-Ogly:2025}, we conducted a thorough examination of all possible ordering configurations along the $x$ axis and $y$ axis. The most intricate and compelling cases of magnetism formation in a decorated square lattice emerge when the system encounters competing exchange interactions in two directions of the lattice concurrently. The following discussion will henceforth examine such a scenario with particularities. Consequently, a systematic examination of all possible indications of exchange interactions and parities of the decoration multiplicity parameter yields a total of sixteen cases in which competing exchange interactions arise simultaneously along two directions of the decorated square lattice. The following cases will be examined in this study.

1. The first case is characterized by the fact that all exchange interactions in the spin system are antiferromagnetic, both direct and decorating exchange interactions
\begin{equation}
J_{x}<0,\quad J_{y}<0,\quad J_{dx}<0,\quad J_{dy}<0.\label{eq:JJd:mm}
\end{equation}
However, the competition between exchange interactions arises here only when the decoration multiplicity is odd in both directions of the lattice simultaneously, that is, when 
\[
d_{x}=2k_{x}+1,\quad k_{x}\in\mathbb{Z};\quad d_{y}=2k_{y}+1,\quad k_{y}\in\mathbb{Z}.
\]
In this case, the condition (\ref{eq:CJ:o}) is satisfied simultaneously along both directions of the lattice.

2. The second case is characterized by the fact that the direct exchange interactions are ferromagnetic, while the decorating exchange interactions are antiferromagnetic
\begin{equation}
J_{x}>0,\quad J_{y}>0,\quad J_{dx}<0,\quad J_{dy}<0,\label{eq:JJd:pm}
\end{equation}
for even decoration multiplicity, in two directions of the lattice at once
\[
d_{x}=2k_{x},\quad k_{x}\in\mathbb{Z};\quad d_{y}=2k_{y},\quad k_{y}\in\mathbb{Z}
\]
which give rise to competitive exchange interactions within the spin system. In this case, the condition (\ref{eq:CJ:e}) is satisfied along two lattice directions.

3. The third and fourth cases of competing exchange interactions arise in the situation opposite to that described in the previous point, where the direct exchange interactions are antiferromagnetic, and the decorating exchange interactions are ferromagnetic 
\begin{equation}
J_{x}<0,\quad J_{y}<0,\quad J_{dx}>0,\quad J_{dy}>0,\label{eq:JJd:mp}
\end{equation}
both for odd
\[
d_{x}=2k_{x}+1,\quad k_{x}\in\mathbb{Z};\quad d_{y}=2k_{y}+1,\quad k_{y}\in\mathbb{Z}
\]
and for even values of the decoration multiplicities
\[
d_{x}=2k_{x},\quad k_{x}\in\mathbb{Z};\quad d_{y}=2k_{y},\quad k_{y}\in\mathbb{Z}
\]
along two lattice directions simultaneously.

Note that in the cases listed above, all parameters are symmetric with respect to the substitution of $x$ for $y$. Below, we list sets of parameters for which the system also experiences competition between exchange interactions, but these are not symmetric with respect to the substitution of $x$ for $y$---that is, they differ in the signs of the direct exchange interactions ($\sign J_{x}\neq\sign J_{y}$) or the decorating exchange interactions ($\sign J_{dx}\neq\sign J_{dy}$), as well as by the parity of the decoration multiplicities ($d_{x}\modulo2\neq d_{y}\modulo2$) along different lattice directions.

4. The fifth and sixth cases arise when the relations of exchange interactions is (\ref{eq:JJd:mp}), but with different parity of the decoration multiplicities along the $x$ and $y$ directions of the lattice. Both a first option 
\[
d_{x}=2k_{x}+1,\quad k_{x}\in\mathbb{Z};\quad d_{y}=2k_{y},\quad k_{y}\in\mathbb{Z},
\]
and a second option involving a set of the decoration multiplicities
\[
d_{x}=2k_{x},\quad k_{x}\in\mathbb{Z};\quad d_{y}=2k_{y}+1,\quad k_{y}\in\mathbb{Z}
\]
are possible. It can be seen that these two presented variants are not symmetric, but transform into one another by swapping $x$ and $y$.

5. The seventh and eighth cases are characterized by the fact that the direct exchange interactions are antiferromagnetic, while the decorating exchange interactions differ: ferromagnetic and antiferromagnetic, i.e., 
\[
J_{x}<0,\quad J_{y}<0,\quad J_{dx}>0,\quad J_{dy}<0,
\]
for odd decoration multiplicities along two lattice directions 
\[
d_{x}=2k_{x}+1,\quad k_{x}\in\mathbb{Z};\quad d_{y}=2k_{y}+1,\quad k_{y}\in\mathbb{Z}
\]
or even decoration multiplicities along the $x$ direction and odd decoration multiplicities along the $y$ direction of the lattice
\[
d_{x}=2k_{x},\quad k_{x}\in\mathbb{Z};\quad d_{y}=2k_{y}+1,\quad k_{y}\in\mathbb{Z}.
\]

6. The ninth and tenth cases are characterized by the fact that the direct exchange interactions are antiferromagnetic, while the decorating exchange interactions differ: antiferromagnetic and ferromagnetic, i.e., 
\[
J_{x}<0,\quad J_{y}<0,\quad J_{dx}<0,\quad J_{dy}>0,
\]
for odd decoration multiplicities along two lattice directions
\[
d_{x}=2k_{x}+1,\quad k_{x}\in\mathbb{Z};\quad d_{y}=2k_{y}+1,\quad k_{y}\in\mathbb{Z}
\]
or for odd decoration multiplicities along the $x$ direction and even decoration multiplicities along the $y$ direction of the lattice
\[
d_{x}=2k_{x}+1,\quad k_{x}\in\mathbb{Z};\quad d_{y}=2k_{y},\quad k_{y}\in\mathbb{Z}.
\]
It can be seen that the parameters in points 5 and 6 are not symmetric, but transform into one another by swapping $x$ and $y$.

7. The eleventh case is characterized by the fact that the decorating exchange interactions are antiferromagnetic, while the direct exchange interactions vary: they are either ferromagnetic or antiferromagnetic, that is, 
\[
J_{x}>0,\quad J_{y}<0,\quad J_{dx}<0,\quad J_{dy}<0,
\]
for even decoration multiplicities along the $x$ direction and odd decoration multiplicities along the $y$ direction of the lattice
\[
d_{x}=2k_{x},\quad k_{x}\in\mathbb{Z};\quad d_{y}=2k_{y}+1,\quad k_{y}\in\mathbb{Z}.
\]

8. The twelfth case is characterized by the fact that the decorating exchange interactions are antiferromagnetic, while the direct exchange interactions are of different types: ferromagnetic and antiferromagnetic, i.e., 
\[
J_{x}<0,\quad J_{y}>0,\quad J_{dx}<0,\quad J_{dy}<0,
\]
for odd decoration multiplicities along the $x$ direction and even decoration multiplicities along the $y$ direction of the lattice
\[
d_{x}=2k_{x}+1,\quad k_{x}\in\mathbb{Z};\quad d_{y}=2k_{y},\quad k_{y}\in\mathbb{Z}.
\]
It can be seen that the parameters in points 7 and 8 are not symmetric, but transform into one another by swapping $x$ and $y$.

9. The thirteenth and fourteenth cases are characterized by the fact that the direct and decorating exchange interactions are different: ferromagnetic and antiferromagnetic direct exchange interactions, as well as antiferromagnetic and ferromagnetic decorating exchange interactions, i.e., 
\[
J_{x}>0,\quad J_{y}<0,\quad J_{dx}<0,\quad J_{dy}>0,
\]
for even decoration multiplicities along the $x$ direction and odd decoration multiplicities along the $y$ direction of the lattice
\[
d_{x}=2k_{x},\quad k_{x}\in\mathbb{Z};\quad d_{y}=2k_{y}+1,\quad k_{y}\in\mathbb{Z}
\]
or for even decoration multiplicities along both directions of the lattice
\[
d_{x}=2k_{x},\quad k_{x}\in\mathbb{Z};\quad d_{y}=2k_{y},\quad k_{y}\in\mathbb{Z}.
\]

10. The fifteenth and sixteenth cases are distinguished by the fact that the direct and decorating exchange interactions are different: antiferromagnetic and ferromagnetic direct, as well as ferromagnetic and antiferromagnetic decorating exchange interactions, i.e.,
\[
J_{x}<0,\quad J_{y}>0,\quad J_{dx}>0,\quad J_{dy}<0,
\]
for odd decoration multiplicities along the $x$ direction and even decoration multiplicities along the $y$ direction of the lattice
\[
d_{x}=2k_{x}+1,\quad k_{x}\in\mathbb{Z};\quad d_{y}=2k_{y},\quad k_{y}\in\mathbb{Z}
\]
or for even decoration multiplicities along both directions of the lattice
\[
d_{x}=2k_{x},\quad k_{x}\in\mathbb{Z};\quad d_{y}=2k_{y},\quad k_{y}\in\mathbb{Z}.
\]
It can be seen that the parameters in items 9 and 10 are not symmetric, but transform into one another by swapping $x$ and $y$.

\section{Isotropic spin system}
\label{sec:iso}

Let us consider the case of an isotropic Ising system on a decorated square lattice (see the example in Fig.~\ref{fig:01}), that is, a situation where the lattice is decorated identically along two lattice directions, with the direct exchange interaction along the $x$ and $y$ axes being equal to each other and the decorating exchange interaction along the $x$ and $y$ axes also being equal to each other,
\begin{equation}
J_{x}=J_{y}=J,\quad d_{x}=d_{y}=d,\quad J_{dx}=J_{dy}=J_{d}.\label{eq:J:iso}
\end{equation}
In such an isotropic spin system, the competition between exchange interactions along two lattice directions simultaneously is possible only in the cases described above (\ref{eq:JJd:mm}), (\ref{eq:JJd:pm}), and (\ref{eq:JJd:mp}).

\begin{figure}[tbh]
\centering \includegraphics{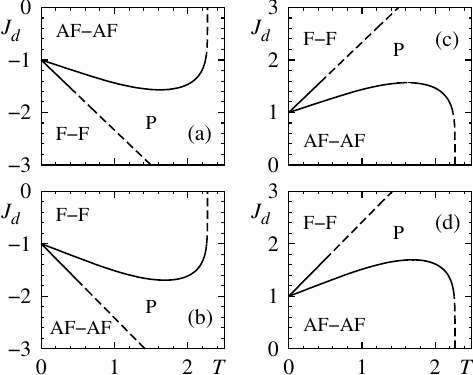}
\caption{Magnetic phase diagrams showing the dependence of the critical temperature $T_{\text{c}}$ on the values of the decorating exchange interaction $J_{d}$ at $J=-1$, $d=5$, $J_{d}<0$ (a), $J=+1$, $d=6$, $J_{d}<0$ (b), $J=-1$, $d=5$, $J_{d}>0$ (c), $J=-1$, $d=6$, $J_{d}>0$ (d). Magnetic ordered phases are labeled in the figure as F--F (ferromagnetic-ferromagnetic) and AF--AF (antiferromagnetic-antiferromagnetic), and the paramagnetic phase as~P}
\label{fig:02}
\end{figure}

For these sets of model parameters, Fig.~\ref{fig:02} shows examples of magnetic phase diagrams (temperature dependence of the decorating exchange interaction) for $|J|=1$, illustrating the behavior of the boundaries of magnetic ordered regions. It can be seen that the magnetic phase diagram contains two regions with magnetic order (ferromagnetic-ferromagnetic (F--F), antiferromagnetic-antiferromagnetic (AF--AF)), as well as a region with a paramagnetic phase (P).

It should be noted that all four magnetic phase diagrams (Fig.~\ref{fig:02}) exhibit a similar structure of the $J_{d}(T)$ dependence. Furthermore, all the plots show that as the temperature increases at 
\begin{equation}
|J_{d}|>|J|\label{eq:R}
\end{equation}
a region appears on the diagram where the spin system undergoes multiple magnetic phase transitions, i.e., the region of magnetic reentrancy. (The intervals of magnetic reentrancy are marked in Fig.~\ref{fig:02} by solid lines representing the boundaries of the magnetic phase regions, while the intervals in which magnetic reentrancy does not occur are marked by dashed lines representing the phase boundaries.) The magnetic phase diagrams show that the phenomenon of magnetic reentrancy does not occur throughout the entire range of model parameter relations in which competing exchange interactions (\ref{eq:CJ:o}) or (\ref{eq:CJ:e}) are active, but only in the range of exchange interaction parameter relations (\ref{eq:R}).

Thus, in the region of magnetic reentrancy, the spin system is magnetically ordered at zero temperature and, as the temperature rises, undergoes a magnetic transition from the magnetically ordered phase to the paramagnetic phase. With a further increase in temperature, the paramagnetic phase transitions restore to the magnetic phase, and with a subsequent increase in temperature, the magnetic phase breaks down, finally forming the paramagnetic region on the phase diagram.

To illustrate this magnetic behavior of a decorated spin system in greater detail, let us consider the following set of parameters for the model 
\begin{equation}
J=-1,\quad d=6,\quad J_{d}>0.\label{eq:par}
\end{equation}
It should also be noted here that the parameters are chosen such that a regime of competing exchange interactions arises along both directions of the decorated square lattice, in accordance with the condition (\ref{eq:CJ:e}). This corresponds to the case described above (\ref{eq:JJd:mp}).

\begin{figure}[tbh]
\centering \includegraphics{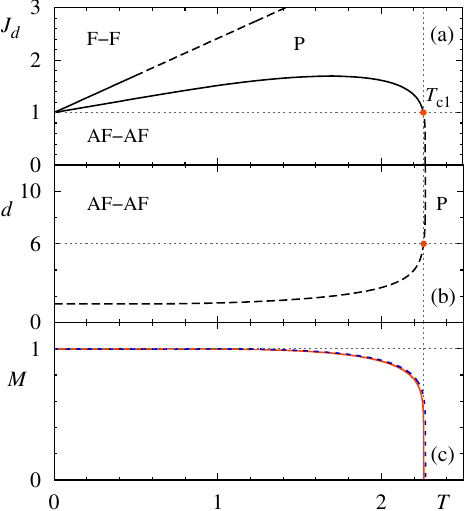}
\caption{Magnetic phase diagrams showing the dependence of the critical temperature $T_{\text{c}}$ on the decorating exchange interaction $J_{d}$ (a) and on the decoration multiplicity $d$ (b) of the lattice for the set of parameters (\ref{eq:par:10}). The red dot indicates the critical temperature $T_{\text{c1}}\approx2.258$ at $J_{d}=+1$ (a) and at $d=6$ (b). The temperature dependence of the spontaneous magnetization (c) for the parameters (\ref{eq:par:10}) is shown by the red solid line, while the Onsager's spontaneous magnetization (\ref{eq:M:O}) with $T_{\text{c}}\approx2.269$ is shown by the blue dashed line}
\label{fig:03}
\end{figure}

\begin{figure}[tbh]
\centering \includegraphics{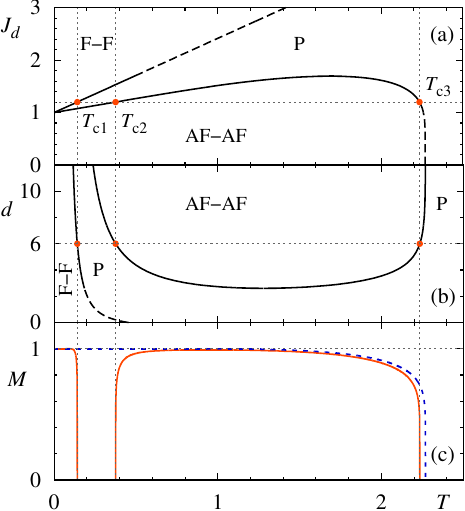}
\caption{Magnetic phase diagrams showing the dependence of the critical temperature $T_{\text{c}}$ on the decorating exchange interaction $J_{d}$ (a) and on the decoration multiplicity $d$ (b) of the lattice for the set of parameters (\ref{eq:par:12}). The red dots indicate the critical temperatures $T_{\text{c1}}\approx0.141$, $T_{\text{c2}}\approx0.376$, $T_{\text{c3}}\approx2.234$ at $J_{d}=+1.2$ (a) and at $d=6$ (b). The temperature dependence of the spontaneous magnetization (c) for the parameters (\ref{eq:par:12}) is shown by the red solid line, while the Onsager's spontaneous magnetization (\ref{eq:M:O}) with $T_{\text{c}}\approx2.269$ is shown by the blue dashed line}
\label{fig:04}
\end{figure}

Note that for the set of model parameters (\ref{eq:par}), under the condition $|J_{d}|<|J|$, the spin system undergoes a single magnetic phase transition. This can be seen in Fig.~\ref{fig:03}, plotted for the parameters 
\begin{equation}
J=-1,\quad d=6,\quad J_{d}=+1.\label{eq:par:10}
\end{equation}
On the other hand, the spin system for the set of parameters (\ref{eq:par}) under the condition (\ref{eq:R}) can already undergo three magnetic phase transitions, as shown in Fig.~\ref{fig:04} for the set of parameters
\begin{equation}
J=-1,\quad d=6,\quad J_{d}=+1.2.\label{eq:par:12}
\end{equation}
That is, the spin system is already in the regime of magnetic reentrancy.

Figures \ref{fig:03}(a) and \ref{fig:04}(a) show magnetic phase diagrams illustrating how the critical temperature of the system depends on the decorating exchange interaction parameter ($J_{d}$). The horizontal dashed lines on the graphs show the parameter $J_{d}$ from the sets (\ref{eq:par:10}) and (\ref{eq:par:12}), respectively. It can be seen that these magnetic phase diagrams contain two regions with magnetic order (ferromagnetic-ferromagnetic (F--F) and antiferromagnetic-antiferromagnetic (AF--AF)), as well as a region with a paramagnetic phase (P).

Figures \ref{fig:03}(b) and \ref{fig:04}(b) show magnetic phase diagrams depicting the dependence of the critical temperature on the decoration multiplicity ($d$). The horizontal dashed line on the graphs represents the parameter $d=6$, defined in the sets (\ref{eq:par:10}) and (\ref{eq:par:12}). As in the previous case, the intervals of magnetic reentrancy are marked by solid lines representing the boundaries of the magnetic phase regions, while the intervals in which magnetic reentrancy does not occur are marked by dashed lines representing the magnetic phase boundaries.

Figures \ref{fig:03}(c) and \ref{fig:04}(c) also show the temperature dependence of the spontaneous magnetization of the spin system under study as a solid line, as well as the Onsager's spontaneous magnetization (\ref{eq:M:O}) as a dashed line.

Figure \ref{fig:03} shows that at $J_{d}=+1$ (which corresponds to the set of parameters (\ref{eq:par:10})) a magnetic phase with AF--AF ordering forms at low temperatures ($0<T<T_{\text{c1}}$), which breaks down as the temperature increases ($T>T_{\text{c1}}$), forming a paramagnetic phase. In this case, the spin system undergoes only one magnetic phase transition at temperature $T_{\text{c1}}$.

Figure \ref{fig:04} shows that at $J_{d}=+1.2$ (which corresponds to the set of parameters (\ref{eq:par:12})) a magnetic phase with F--F ordering forms at low temperatures ($0<T<T_{\text{c1}}$), which breaks down as the temperature increases ($T_{\text{c1}}<T<T_{\text{c2}}$), forming a paramagnetic phase. With a further increase in temperature ($T_{\text{c2}}<T<T_{\text{c3}}$), a magnetic phase with AF--AF ordering reenters in the system, which subsequently breaks down as the temperature rises ($T>T_{\text{c3}}$), finally forming a paramagnetic region. That is, in this case, the spin system undergoes three magnetic phase transitions at temperatures $T_{\text{c1}}$, $T_{\text{c2}}$, and $T_{\text{c3}}$.

Thus, it can be seen that, as the temperature changes, a decorated spin system with a specific set of parameters can enter a state in which magnetic and non-magnetic regions alternate, that is, a regime of magnetic reentrancy. In this case, an isotropic Ising system on a decorated square lattice may undergo one or three magnetic phase transitions.

\section{Anisotropic spin system}

In the event that the model parameters differ along the $x$ and $y$ directions, the spin system becomes anisotropic. There are an infinite number of possible combinations of model parameters that satisfy, among others, the following relations: (\ref{eq:JJd:mm}), (\ref{eq:JJd:pm}), (\ref{eq:JJd:mp}), as well as the others described in Section \ref{sec:conc}, under which magnetic reentrancy may occur.

\begin{figure}[tbh]
\centering \includegraphics{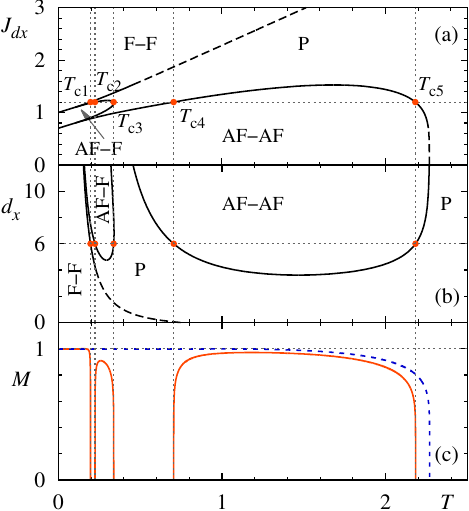}
\caption{Magnetic phase diagrams showing the dependence of the critical temperature $T_{\text{c}}$ on the decorating exchange interaction $J_{dx}$ ($J_{dy}=J_{dx}+0.3$) (a) and the decoration multiplicity $d_{x}$ ($d_{y}=d_{x}$) (b) of the lattice for the set of parameters (\ref{eq:par:JD}). The red dots denote the critical temperatures $T_{\text{c1}}\approx0.197$, $T_{\text{c2}}\approx0.224$, $T_{\text{c3}}\approx0.338$, $T_{\text{c4}}\approx0.705$, $T_{\text{c5}}\approx2.183$ at $J_{dx}=+1.2$ ($J_{dy}=+1.5$) (a) and at $d_{x}=d_{y}=6$ (b). The temperature dependence of the spontaneous magnetization (c) for the parameters (\ref{eq:par:JD}) is shown by the red solid line, and the Onsager's spontaneous magnetization (\ref{eq:M:O}) with $T_{\text{c}}\approx2.269$ is shown by the blue dashed line}
\label{fig:05}
\end{figure}

To illustrate the behavior of an anisotropic decorated spin system, we will lean on the results from the previous Section \ref{sec:iso}, modifying only the decorating exchange interactions in the set of parameters (\ref{eq:par}) so that they are unequal to each other
\[
J_{x}=J_{y}=-1,\quad d_{x}=d_{y}=6,
\]
\begin{equation}
J_{dx}=+1.2,\quad J_{dy}=+1.5.\label{eq:par:JD}
\end{equation}
It should be noted here that the parameters are chosen such that, along each of the $x$ and $y$ directions, direct and decorating exchange interactions compete with one another, in accordance with condition (\ref{eq:CJ:e}). In this case, the relation between direct and decorating exchange interactions satisfies condition (\ref{eq:R}), and the relation between decorating exchange interactions satisfies the criterion
\begin{equation}
|J_{dx}|\neq|J_{dy}|.\label{eq:R:JD}
\end{equation}
It is evident that the difference between the parameters $J_{dx}$ and $J_{dy}$ causes the spin system to exhibit different conditions for the formation of magnetic regions along different directions of the decorated lattice, which leads to a change in the magnetic phase diagram, including the formation of as many as five magnetic phase transitions (as shown in Fig.~\ref{fig:05}).

Note that the structure of the sets of graphs in Fig.~\ref{fig:05} and the following figures is similar to that of the graphs in Figs.~\ref{fig:03} and \ref{fig:04}. The magnetic phase diagram shown in Fig.~\ref{fig:05}, unlike the previous two graphs, already contains three regions with magnetic order (ferromagnetic-ferromagnetic (F--F), antiferromagnetic-ferromagnetic (AF--F), antiferromagnetic-antiferromagnetic (AF--AF)), as well as a region with a paramagnetic phase (P).

Figure \ref{fig:05} shows that at $J_{dx}=+1.2$ (which corresponds to the set of parameters (\ref{eq:par:JD})) at low temperatures ($0<T<T_{\text{c1}}$), a phase with F--F ordering forms, which, as the temperature increases ($T_{\text{c1}}<T<T_{\text{c2}}$), breaks down, forming a paramagnetic phase. With a further increase in temperature ($T_{\text{c2}}<T<T_{\text{c3}}$), a phase with magnetic AF--F ordering emerges in the system, which subsequently breaks down in the temperature range $T_{\text{c3}}<T<T_{\text{c4}}$. A subsequent increase in temperature ($T_{\text{c4}}<T<T_{\text{c5}}$) leads to the formation of a region of magnetic AF--AF ordering, which breaks down at $T>T_{\text{c5}}$, finally forming a paramagnetic region. In this case, the spin system undergoes five magnetic phase transitions at temperatures $T_{\text{c1}}$, $T_{\text{c2}}$, $T_{\text{c3}}$, $T_{\text{c4}}$, and $T_{\text{c5}}$.

\begin{figure}[tbh]
\centering \includegraphics{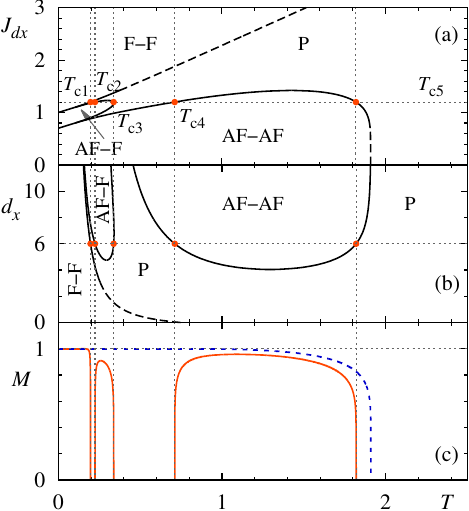}
\caption{Magnetic phase diagrams showing the dependence of the critical temperature $T_{\text{c}}$ on the decorating exchange interaction value $J_{dx}$ $(J_{dy}=J_{dx}$) (a) and on the decoration multiplicity $d_{x}$ ($d_{y}=d_{x}$) (b) of the lattice for the set of parameters (\ref{eq:par:J}). The red dots denote the critical temperatures $T_{\text{c1}}\approx0.197$, $T_{\text{c2}}\approx0.224$, $T_{\text{c3}}\approx0.338$, $T_{\text{c4}}\approx0.712$, $T_{\text{c5}}\approx1.820$ at $J_{dx}=+1.2$ ($J_{dy}=J_{dx}$) (a) and at $d_{x}=d_{y}=6$ (b). The temperature dependence of the spontaneous magnetization (c) for the parameters (\ref{eq:par:J}) is shown by the red solid line, and the Onsager's spontaneous magnetization (\ref{eq:M:O}) with $T_{\text{c}}\approx1.910$ is shown by the blue dashed line}
\label{fig:06}
\end{figure}

On the other hand, we can change the relation between direct exchange interactions along the $x$ and $y$ axes, rather than those of decorating exchange interactions; for example,
\[
J_{x}=-1,\quad J_{y}=-0.7,
\]
\begin{equation}
d_{x}=d_{y}=6,\quad J_{dx}=J_{dy}=+1.2.\label{eq:par:J}
\end{equation}
With this set of parameters, we have obtained an anisotropic spin system. As before, the parameters are chosen such that, along each of the $x$ and $y$ directions, direct and decorating exchange interactions compete, in accordance with the condition ((\ref{eq:CJ:o})). In this case, the relation between direct and decorating exchange interactions satisfies condition (\ref{eq:R}), and the relation between direct exchange interactions satisfies the criterion
\begin{equation}
|J_{x}|\neq|J_{y}|.\label{eq:R:J}
\end{equation}
It is evident that the differences in the parameters $J_{x}$ and $J_{y}$ result in the spin system having different conditions for the formation of magnetic regions along different directions of the decorated lattice, which leads to a change in the magnetic phase diagram, including the formation of five magnetic phase transitions (as shown in Fig.~\ref{fig:06}).

The behavior of the magnetic phases in the diagrams shown in Fig.~\ref{fig:06} is similar to that described above for Fig.~\ref{fig:05}; only the critical temperatures differ. As in the previous case, the system exhibits magnetic reentrancy and possesses five magnetic transitions.

\begin{figure}[tbh]
\centering \includegraphics{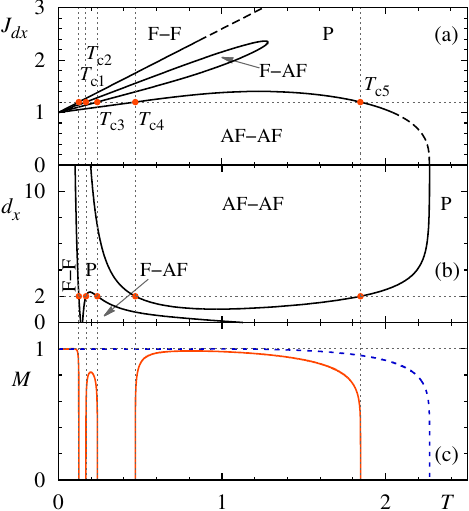}
\caption{Magnetic phase diagrams showing the dependence of the critical temperature $T_{\text{c}}$ on the decorating exchange interaction value $J_{dx}$ ($J_{dy}=J_{dx}$) (a) and on the decoration multiplicity $d_{x}$ $(d_{y}=d_{x}+16$) (b) of the lattice for the set of parameters (\ref{eq:par:D}). The red dots denote the critical temperatures $T_{\text{c1}}\approx0.125$, $T_{\text{c2}}\approx0.169$, $T_{\text{c3}}\approx0.238$, $T_{\text{c4}}\approx0.471$, $T_{\text{c5}}\approx1.847$ at $J_{dx}=+1.2$ ($J_{dy}=J_{dx}$) (a) and at $d_{x}=2$ ($d_{y}=18$) (b). The temperature dependence of the spontaneous magnetization (c) for the parameters (\ref{eq:par:D}) is shown by the red solid line, and the Onsager's spontaneous magnetization (\ref{eq:M:O}) with $T_{\text{c}}\approx2.269$ is shown by the blue dashed line}
\label{fig:07}
\end{figure}

It is also possible to change not only the relation between the values of exchange interaction, but also the lattice decoration multiplicities, for example, 
\[
J_{x}=J_{y}=-1,\quad d_{x}=2,\quad d_{y}=18,
\]
\begin{equation}
J_{dx}=J_{dy}=+1.2.\label{eq:par:D}
\end{equation}
With this set of parameters, we also obtained an anisotropic spin system. As before, the parameters are chosen such that, along each of the $x$ and $y$ directions, direct and decorating exchange interactions compete, in accordance with the condition (\ref{eq:CJ:e}). In this case, the relation between direct and decorating exchange interactions satisfies the condition (\ref{eq:R}), and the decoration multiplicities satisfy the criterion
\begin{equation}
|d_{x}-d_{y}|=k,\quad k>0,\quad k\in\mathbb{Z}.\label{eq:R:D}
\end{equation}
It is evident that the differences in the parameters $d_{x}$ and $d_{y}$ result in the spin system having different conditions for the formation of magnetic regions along different directions of the decorated lattice. This leads to a change in the magnetic phase diagram, including the formation of five magnetic phase transitions (as shown in Fig.~\ref{fig:07}).

The behavior of the magnetic phases in the diagrams shown in Fig.~\ref{fig:07} is similar to that described above (Figs.~\ref{fig:05} and \ref{fig:06}); the only difference here is in the critical temperatures. As in the previous case, the system exhibits magnetic reentrancy and also has five magnetic phase transitions.

In this section, we have considered cases that differ from the isotropic case (Section \ref{sec:iso}), where only individual pairs of parameters along the $x$ and $y$ directions differ. Of course, more complicated combinations of model parameter sets are possible, which would also alter the behavior of magnetic phases in the phase diagram. However, this study and our work \citep{Kassan-Ogly:2025} suggest that the number of magnetic phase transitions will remain the same: either one, three, or five.

\section{On the calculation of critical temperatures of magnetic phase transitions}

Let us turn to the crucial issue of determining the values of critical temperatures and the number of magnetic phase transitions. This task takes on particular significance because variations in model parameters often lead to situations involving fundamental difficulties in identifying the characteristics of the transition. The problem is complicated by the fact that in the immediate vicinity of critical points, the behavior of the system thermodynamic and magnetic characteristics undergoes such abrupt changes that the direct and precise determination of transition temperatures becomes a non-trivial computational problem. Moreover, as established in the previous section, the spin systems under consideration may exhibit several successive magnetic phase transitions, including transition points that are quite close to one another, as well as transitions in the region of extremely low temperatures, which requires the use of specialized analytical and numerical approaches. We also note that the calculation of thermodynamic functions near extremely low temperatures faces fundamental limitations on the applicability of the computational methods and approaches used, which is a well-known problem and is associated with difficulties in achieving high reliability of results. The correct and accurate determination of the magnetic phase transition temperatures of such magnetics is extremely important for the adequate description, understanding, and prediction of the magnetic behavior of the spin systems under study.

The search for critical temperatures involves identifying the points at which spontaneous magnetization appears and disappears. However, in practice, the temperature of the magnetic phase transition is often determined by the position of the logarithmic divergence of the temperature dependence of heat capacity. This is because no expression for the spontaneous magnetization of the system has been found for all problems (models and lattices).

It is known that, in the theory of phase transitions, the Onsager's logarithmic divergence at critical temperatures of heat capacity \citep{Onsager:1944},
\begin{equation}
C=2\frac{T}{\lambda}\frac{\partial\lambda}{\partial T}+\frac{T^{2}}{\lambda}\frac{\partial^{2}\lambda}{\partial T^{2}}-\frac{T^{2}}{\lambda^{2}}\left(\frac{\partial\lambda}{\partial T}\right)^{2}\label{eq:C}
\end{equation}
is directly related to the divergence of the subintegral logarithm
(\ref{eq:Lf}),
\begin{equation}
\Lambda(\phi_{x},\phi_{y})=C_{x}C_{y}-S_{x}D_{y}\cos\phi_{x}-D_{x}S_{y}\cos\phi_{y},\label{eq:Lf:}
\end{equation}
at four points (\ref{eq:Lf:a}). Based on this, it is possible to construct a procedure for finding critical temperatures by computing the roots of the function (\ref{eq:Lf:}),
\begin{equation}
\Lambda(\phi_{x},\phi_{y})=0,\label{eq:Lf:0}
\end{equation}
at the corresponding four points (\ref{eq:Lf:a}), that is, the following four functions
\[
\Lambda(0,0)=C_{x}C_{y}-S_{x}D_{y}-D_{x}S_{y},
\]
\[
\Lambda(0,\pi)=C_{x}C_{y}-S_{x}D_{y}+D_{x}S_{y},
\]
\[
\Lambda(\pi,0)=C_{x}C_{y}+S_{x}D_{y}-D_{x}S_{y},
\]
\begin{equation}
\Lambda(\pi,\pi)=C_{x}C_{y}+S_{x}D_{y}+D_{x}S_{y}.\label{eq:Lf:p}
\end{equation}

To implement this computational procedure, the following variable substitution should be made in the function arguments (\ref{eq:Lf:p}):
\begin{equation}
t=\exp\left(\frac{1}{rT}\right).\label{eq:x}
\end{equation}
Here, the coefficient $r$ is chosen in accordance with the specified model parameters so that the values of $rJ_{i}$ and $rJ_{di}$ are integers. In this case, the functions $C_{i}$ (\ref{eq:Ci}), $S_{i}$ (\ref{eq:Si}), and $D_{i}$ (\ref{eq:Di}) are rewritten taking into account the following substitutions:
\[
e^{2\frac{J_{i}}{T}}=t^{2J_{i}r},
\]
\[
\cosh\frac{J_{di}}{T}=\frac{e^{\frac{J_{di}}{T}}+e^{-\frac{J_{di}}{T}}}{2}=\frac{t^{J_{di}r}+t^{-J_{di}r}}{2},
\]
\[
\sinh\frac{J_{di}}{T}=\frac{e^{\frac{J_{di}}{T}}-e^{-\frac{J_{di}}{T}}}{2}=\frac{t^{J_{di}r}-t^{-J_{di}r}}{2}.
\]
Thus, by expressing all the functions (\ref{eq:Lf:p}) in terms of the new variable $t$ (\ref{eq:x}), we transform the problem of analyzing phase transitions into a search for the roots of these four algebraic power equations.

The root-finding procedure ($t_{i}$) for the equations (\ref{eq:Lf:0}) described above allows us to calculate the number and values of critical temperatures in a spin system. The temperature values are expressed in terms of the function (\ref{eq:x}) as 
\begin{equation}
T_{\text{c}i}=\frac{1}{r\ln t_{i}},\quad t_{i}\in\mathbb{R},\quad t_{i}>1,\label{eq:Tx}
\end{equation}
where $t_{i}$ are the real roots of the equations, greater than unity. This condition corresponds to physically realizable phase transitions in which the temperature has a positive value. 

Note that, depending on the chosen model parameters and the coefficient $r$, the degree of the equations to be solved may be quite high; luckily, the number of roots constrained by the condition (\ref{eq:Tx}) always coincides with the number of phase transitions. This approach, based on rigorous mathematical methods, provides the most reliable identification and characterization of the positions of phase transition points.

\begin{figure}[tbh]
\centering \includegraphics{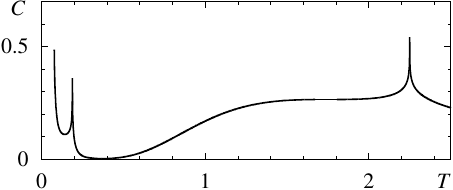}
\caption{Temperature dependence of heat capacity (\ref{eq:C}) on model parameters (\ref{eq:parX})}
\label{fig:08}
\end{figure}

For the sake of argument, let us consider the following example and calculate the temperature dependence of the heat capacity of an isotropic decorated square lattice for the following set of model parameters:
\begin{equation}
J=-1,\quad d=6,\quad J_{d}=+1.1.\label{eq:parX}
\end{equation}
Figure \ref{fig:08} shows that at temperatures $T>0.1$, the system undergoes two magnetic phase transitions at $T_{\text{c}}\approx0.188$ and $T_{\text{c}}\approx2.249$, which can be determined from the positions of the lambda-shaped heat capacity peaks. However, calculations performed at low temperatures ($T<0.1$) lead to physically incorrect results (see the low-temperature region in Fig.~\ref{fig:08}). This fact indicates that it is impossible to obtain reliable values of the heat capacity in the region of extremely low temperatures, as noted above.

To implement our approach, we must set the coefficient $r$ in the variable (\ref{eq:x}) such that the desired expressions take the form of a polynomial in the variable $t$, where, regardless of the model parameters, the degrees of the polynomials are integers. Therefore, for a given set of parameters (\ref{eq:parX}), it is convenient to choose the coefficient of the variable (\ref{eq:x}) to be $r=10$. In this case, the degree of the polynomial to be solved, constructed for the equations (\ref{eq:Lf:p}), will be $268$. Despite the high degree of the equation to be solved, in this case there are only three roots of the equations (\ref{eq:Lf:p}) that satisfy the condition (\ref{eq:Tx}), which are 
\begin{equation}
t_{1}\approx4.111,\quad t_{2}\approx1.703,\quad t_{3}\approx1.045.\label{eq:x3}
\end{equation}
The first two roots (\ref{eq:x3}) are solutions to the equation $\Lambda(0,0)=0$, and the last root is a solution to $\Lambda(\pi,\pi)=0$. The remaining $265$ roots are either complex or real, but less than one. From this, using the condition (\ref{eq:Tx}), we can calculate all the critical temperatures, which are equal to
\[
T_{\text{c}1}\approx0.071,\quad T_{\text{c}2}\approx0.188,\quad T_{\text{c}3}\approx2.249.
\]
Thus, we obtained the exact number of critical temperatures and their values. As a result, we found not two, but three temperatures of magnetic phase transitions (see Fig.~\ref{fig:08}).

The proposed method demonstrates how algebraic calculations of the roots of four equations allow us to precisely determine the number and values of the temperatures of magnetic phase transitions, escaping potential errors in determining critical temperatures. With this approach, we are not dependent on the computational difficulties of thermodynamic functions, which significantly simplifies the analysis of multiple phase transitions, including those with closely spaced critical temperatures, especially at low and extremely low temperatures.

\section*{Conclusion}

In this paper, using an exact analytical solution to the Ising model on a decorated square lattice with an arbitrary number of decorating spins along two lattice directions, we demonstrated the possibility of describing the phenomenon of multiple magnetic phase transitions as the temperature increases, namely magnetic reentrancy.

Spin systems with magnetic reentrancy exhibit a complicated magnetic phase diagram, which is a natural consequence of the multi-parameter dependence of the model used on a decorated square lattice. Along with direct and decorating exchange interactions, which characterize the energy parameters of the spin couplings along two directions, and two parameters determining the decoration multiplicity of the square lattice in its various directions, temperature also acts as a crucial factor shaping the magnetic order. In such complicated system of interrelated parameters, competition may arise between different magnetic states, leading to a sequential, multiple change in the dominant type of magnetic order and the paramagnetic phase as the temperature changes. This is precisely how the phenomenon of magnetic reentrancy is formed.

To understand the conditions under which the phenomenon of magnetic reentrancy arises, this study calculated the magnetic phase diagrams and temperature functions of the spontaneous magnetization of an Ising spin system on a decorated square lattice for various model parameters. A regime of model parameter relations was identified in which multiple magnetic phase transitions occur in such a decorated spin system. The behavior of magnetic reentrancy as the model parameters vary is described.

It has been established that the occurrence of the magnetic reentrancy phenomenon is due to the presence of competing exchange interactions in the system. This phenomenon arises when the absolute value of the decorator exchange interaction exceeds that of the direct exchange interaction. It is shown that, when these conditions are present simultaneously along two lattice directions, one, three, or five magnetic phase transitions can be observed in the system, which indicates the complicated nature of magnetic interactions in the system under study. Moreover, the presence of as many as five such transitions is due to the anisotropy of the lattice, that is, the difference in the absolute values of the decorating or direct exchange interactions, or the difference in the decorating multiplicities along the two lattice directions. It should be noted, however, that in a decorated square lattice, no more than five magnetic phase transitions occur, nor does an even number of magnetic phase transitions.

Another key result of this work is the described method for calculating the temperatures of magnetic phase transitions. This approach allows for the algebraic determination of the number and values of critical temperatures, which has made it possible to circumvent computational difficulties when critical temperatures are close to one another, as well as at extremely low temperatures.

It is precisely this complicated picture of magnetic ordering and, consequently, the rich magnetic phase diagram that accounts for the particular scientific interest in the study of decorated spin systems. We believe that this study will help elucidate the mechanisms and causes of magnetic reentrancy, including contributing to a deeper understanding of the mechanisms underlying this phenomenon in decorated spin systems, as well as stimulating further research into this specific manifestation of magnetism in various crystal lattices.

\section*{Acknowledgments}

The work was carried out within the framework of the state assignment of the Ministry of Science and Higher Education of the Russian Federation for the IMP UB RAS.

\bibliographystyle{apsrev4-2}
\bibliography{sq1dr005002l}

\end{document}